\documentstyle[times,pramana,graphics,epsf,floats,color]{ias}
\begin{document}
\mark{{The effects of next to nearest neighbor hopping on Bose-Einstein
condensation in cubic lattices}{G. K. Chaudhary and R. Ramakumar
}}
\title{The effects of next to nearest neighbor hopping on Bose-Einstein
condensation in cubic lattices}

\author{G. K. Chaudhary and R. Ramakumar}
\address{Department of Physics and Astrophysics, University of Delhi,
Delhi-110007, Delhi, India}
\abstract{
In this paper, we present results of our calculations on
the effects of next to nearest neighbor boson hopping ($t^{\prime}$) energy on 
Bose-Einstein condensation in cubic lattices. We consider both non-interacting 
and repulsively interacting bosons moving in the lowest Bloch band. 
The interacting bosons are studied 
making use of the Bogoliubov method. We find that the bose condensation 
temperature is enhanced with increasing
$t^{\prime}$ for bosons in a simple cubic (sc) lattice and 
decreases for bosons in body-centered cubic (bcc) and 
face-centered cubic (fcc) lattices. We also find that
interaction induced depletion of the condensate is reduced for bosons
in a sc lattice while is enhanced for bosons in bcc and fcc lattices.}

\maketitle
\section{Introduction}
Studies of Bose-Einstein condensation in optical lattices and 
crystalline lattices is an active field of research both in 
atomic\cite{jaksch,greiner,blochdali,yukalov} and
condensed matter 
physics\cite{alex,snoke,kasprzak,balili,nikuni,ruegg,giamarchi}.
In condensed matter physics, there have been extensive studies of bose
condensation of bipolarons\cite{alex}, exctions\cite{snoke}, 
exciton-polaritons\cite{kasprzak,balili}, and magnons\cite{nikuni,ruegg,giamarchi}. 
Studies of bosons in optical lattices may be said to have received 
a boost with the demonstration of bose condensed to 
Mott insulator transition\cite{greiner} predicted in 
theoretical studies\cite{jaksch,fisher,krauth,sheshadri,freericks}
of strongly interacting lattice bosons. 
In the presently available optical lattices, it has been shown\cite{jaksch}
that it is sufficient to include nearest-neighbor (NN) hopping of bosons in the 
kinetic energy part of the Hamiltonian of the system. 
Nevertheless, considering the fast pace of developments in this field,
it may be useful to investigate the effects of the next to nearest-neighbor
(NNN) hopping  on bose condensation in optical and crystalline lattices.
Recently, we presented a study of the lattice symmetry effects on
bose condensation in cubic lattices\cite{ram}. In that work, we had confined
to NN hopping of lattice bosons. In this paper, we extend 
this work including NNN boson hopping. The bosons are considered to be 
of spin-zero and charged (see also the note
in Ref. \onlinecite{neutral}). We would like to emphasize that we are
not exclusively considering bosons in optical lattices. Our calculations
should be considered in the enlarged context including bose condensation
in crystalline lattices.
In the next section, we describe the models and
methods used in our calculations along with a  discussion
of results. The conclusions are given Sec. III.

\section{Bose condensation in cubic lattices with NNN hopping}
{\em Non-interacting bosons}: Consider bosons moving in cubic lattices. 
The energy eigen-functions of a single
boson moving in a periodic optical or crystalline lattice potential are 
Bloch waves\cite{bloch} and 
energy eigen-values form bands. The Hamiltonian of non-interacting 
bosons in an energy band is:
\begin{equation}
H= \sum_{k}[\epsilon(k)-\mu] c^{\dag}_k c_k\, ,
\end {equation}
where $\epsilon(k)$ is the one-boson energy band structure, $k$ is the
boson quasi-momentum, $\mu$ is the chemical potential, and $c^{\dag}_{k}$ 
is a boson creation operator. Within a tight-binding 
approximation scheme\cite{mermin}, including the NN and the NNN Wannier 
functions overlaps, the {\em s}-band structures we consider for cubic 
lattices are:
\begin{eqnarray}
\epsilon_{sc}(k_x,k_y,k_z) =&-&2t\sum_{\mu=x}^{z}\cos(k_{\mu}) \nonumber \\
       &-&2{t^{\prime}}\sum_{\mu=x}^{z}\sum_{\mu \neq \nu,\nu=x}^{z}\cos(k_{\mu}) \cos(k_{\nu})\,,
\end{eqnarray}

\begin {eqnarray}
\epsilon_{bcc}(k_x,k_y,k_z)=&-&8t\prod_{\mu=x}^{z}
            \cos\left(\frac{k_\mu}{2}\right) \nonumber \\
                &-&2{t^{\prime}}\sum_{\mu=x}^{z}\cos(k_\mu)\,,
\end{eqnarray}

\begin{eqnarray}
\epsilon_{fcc}(k_x,k_y,k_z)=&-&2t\sum_{\mu=x}^{z}\sum_{\mu \neq \nu,\nu=x}^{z}
\cos\left(\frac{k_\mu}{2}\right)\cos\left(\frac{k_{\nu}}{2}\right) \nonumber \\
&-&2{t^{\prime}}\sum_{\mu=x}^{z}\cos(k_{\mu})\,,
\end{eqnarray}
where the lattice constant has been set to unity.
Here $t$ is NN boson hopping energy and $t^{\prime}$ is 
NNN boson hopping energy in the lattice.  
\par
The condensation temperature $T_{B}$ for bosons in these bands can be calculated from the boson number equation
\begin{eqnarray}
n = \frac{1}{{N_x}{N_y}{N_z}}\sum_{k_x}\sum_{k_y}\sum_{k_z}\frac{1}{e^{\frac{\epsilon(k_x,k_y,k_z)-\mu}{k_BT}}-1}\,,
\end{eqnarray}
where $N_s=N_x N_y N_z$ is the total number of lattice sites,
$k_B$ is the Boltzmann constant, $T$ is the temperature, and
$n$ is number of bosons per site. We have numerically solved the
bosons number equation (Eq. 5) to obtain the bose condensation temperature
and ground state occupancy.
The results of these calculations for various lattices considered are shown
in Fig. 1. 
\begin{figure}
\resizebox*{4.5in}{4.6in}{\rotatebox{270}{\includegraphics{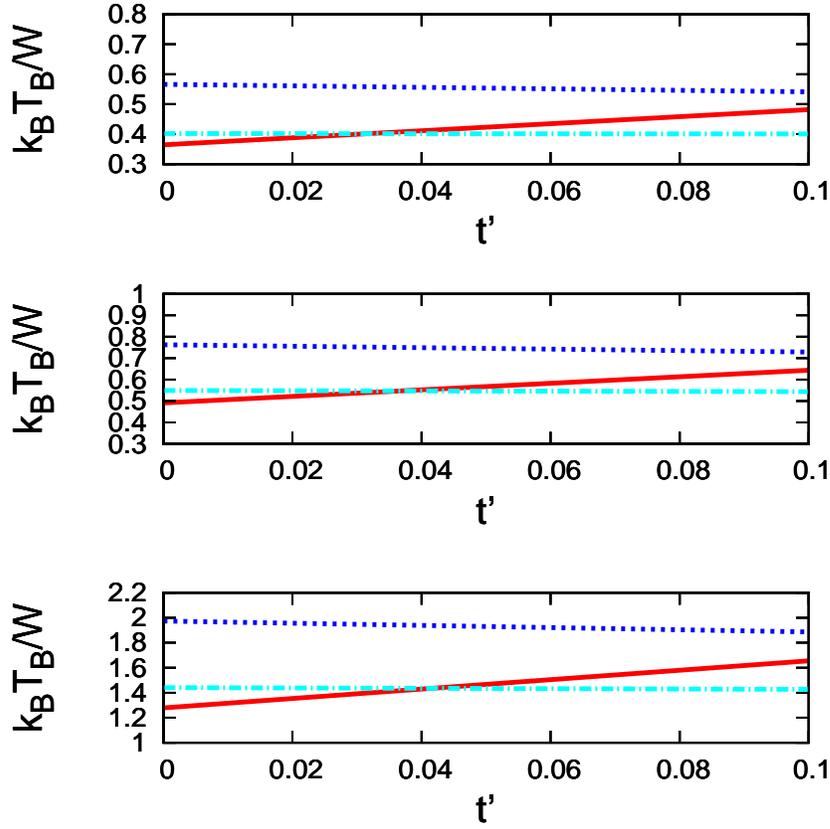}}}
\vspace*{0.5cm}
\caption[]
{
The bose condensation temperature vs NNN hopping $ t^{\prime} $ for noninteracting  bosons in various  cubic lattices. The curves are shown for: 
sc (solid line), bcc (dotted line), and fcc (dash-dot line). These results 
are for n =0.25 (top panel), n=0.4 (middle panel), n=1.5 (bottom panel).In 
this and other figures W is the half-band-width.
}
\label{scaling}
\end{figure}
We find that $t^{\prime}$ increases the bose condensation temperature of bosons in a sc lattice. For bosons in bcc lattice the $T_B$ decreases
with increasing $t^{\prime}$. For bosons in a fcc lattice also increasing 
$t^{\prime}$ more or less leave $T_B$ unaltered. These trends can
be approximately understood in the low boson density limit. In this limit,
states with significant thermal population is close to the bottom of the
energy bands. Now, for a given small boson density, the ratio between
bose condensation temperature and half-band-width is proportional to
$1/(m^{*}W)$, where $m^{*}$ is the boson effective mass. We find
that, in the low density limit,  $T_B/W$ (where $T_B$ is the bose 
condensation temperature and W is the half-band-width) goes 
as: $(t+4t')/3t$ for sc, $(t+t')/(2t+3t')$ for bcc, and $1/3$ for fcc.
On plotting, one can easily see that the $T_B/W$ increases with $t^{\prime}$ in the sc case, decreases slightly for bcc case, and remains 
constant for the fcc case. These trends are consistent with
our numerical results.
\begin{figure}
\resizebox*{4.5in}{4.6in}{\rotatebox{270}{\includegraphics{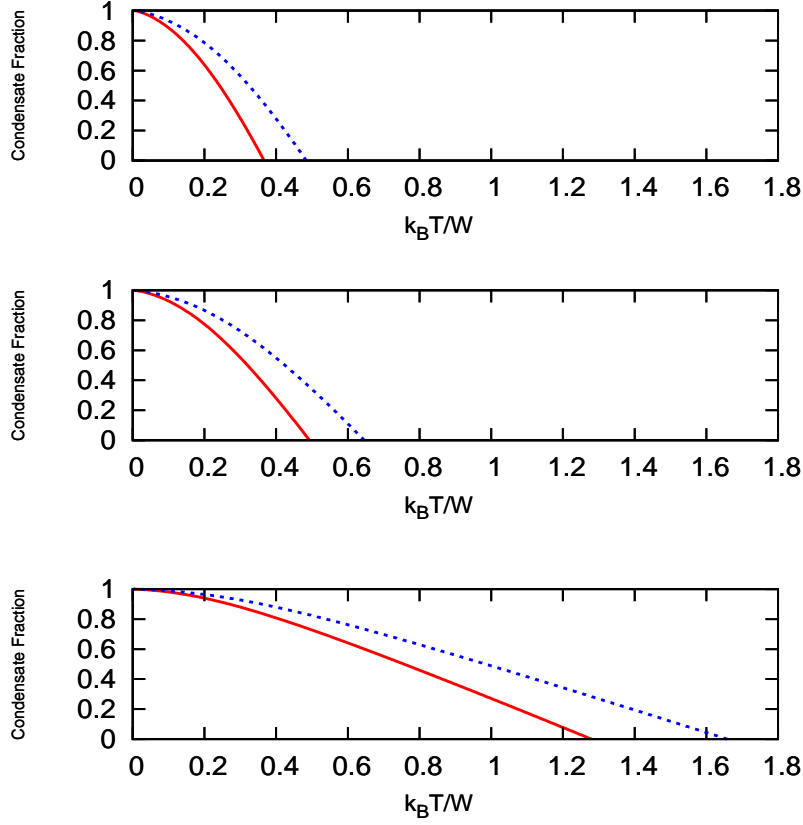}}}
\vspace*{0.5cm}
\caption[]{
The variation of condensate fraction with temperature (T) for bosons in 
a sc lattice with NNN hopping $ t^{\prime}$ : $t^{\prime}=0$ (solid line),  
$t^{\prime}$ = t/10 (dotted line). Here 
  n =0.25 (top panel), n=0.4 (middle panel), n=1.5 (bottom panel).
}
\label{scaling}
\end{figure}
\begin{figure}
\resizebox*{3.1in}{2.5in}{\rotatebox{270}{\includegraphics{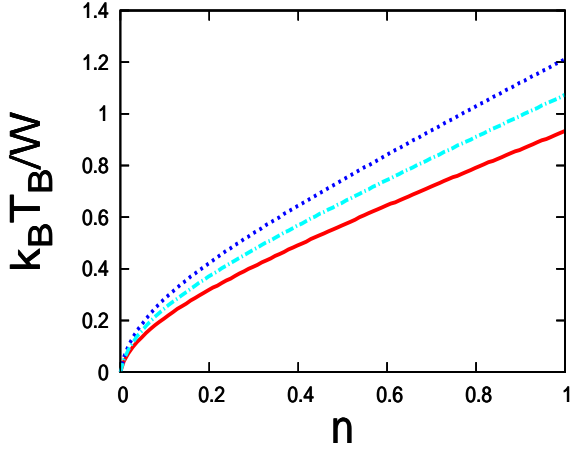}}}
\vspace*{0.5cm}
\caption[]
{
The bose condensation temperature vs n for bosons in a sc lattice with
NNN hopping  $t^{\prime}$ : $t^{\prime}$ =0 (solid line), $t^{\prime}$ = 
t/20 (dash-dot line),
$t^{\prime}$=t/10 (dotted line).
}
\label{scaling}
\end{figure}

The growth of the condensate fraction and the number dependence of $T_B$
for bosons in a sc lattice shown in Fig. 2 and Fig. 3 are similar to that found for the case of $t^{\prime}\,=\,0$\cite{ram}. Similar results are obtained
(not shown) for bosons in bcc and fcc lattices.
\par
{\em Interacting bosons}:  The Hamiltonian of interacting bosons is 
\begin{equation}
H= \sum_{k}[\epsilon(k)-\mu] c^{\dag}_k c_k\ +\frac{U}{2{N_S}}\sum_{k}\sum_{k^{\prime}}\sum_{q}c^{\dag}_{k+q}c^{\dag}_{k^{\prime}-q}c_{k^{\prime}}c_k\,,
\end {equation}
where $U$ is a constant boson-boson repulsive interaction energy.
To treat the effect of interactions we make
use of Bologliubov approach\cite{bogol}
to the interacting bosons system. In this theory, it is assumed
that the ground state of interacting bosons system is a bose
condensate. Since the condensation occurs into the lowest
single particle state (for which ${\bf k} = 0$ in our cases), one 
gets  $<c^{\dag}_{0}c_{0}>\,
\approx\, <c_{0}c^{\dag}_{0}>$.
This allows one to treat the operators $c^{\dag}_{0}$ and $c_{0}$ 
as complex numbers and one gets $<c^{\dag}_{0}>$ = $<c_{0}>$ = $\sqrt{N_0}$.
Here $N_{0}=n_{0}N_{s}$ in which $n_0$ is the boson occupancy per lattice site
in the ${\bf k} = 0$ state. 
The second order interaction
terms are obtained from the substitution:  $c^{\dag}_{0}\rightarrow \sqrt{N_0}+c^{\dag}_{0}$.
On using this approach, the boson number
equation is obtained to be (for details see Ref. 16): 
\begin{eqnarray}
n=n_0+\frac{1}{2{N_s}}\sum^{\prime}_k\left[\left(1+\frac{\xi(k)+U{n_0}}{E(k)}\right)\times \frac{1}{{e^{\frac{E(k)}{k_{B}T}}-1}}\right] \nonumber \\
+\frac{1}{2{N_s}}\sum^{\prime}_k\left[\left(1-\frac{\xi(k)+Un_0}{E(k)}\right)\times\frac{1}{{e^{\frac{-E(k)}{k_{B}T}}-1}}\right]\,,
\end{eqnarray}
where $\xi(k)=\epsilon(k)-
\epsilon_0$, $\epsilon_0$ is the energy of the lowest single particle state,
and  $E(k)=\sqrt{\xi^{2}(k)+2Un_{0}\xi(k)}$.
The primes on the summation signs indicates that the sums excludes the $k=0$ 
state in to which the bosons condense. 
The Bologliubov method used would valid 
so long as the interaction
energy is smaller than the kinetic energy of the bosons.
This approximately translates to $U \leq 2W$. The effect
of increasing interaction (U) is to lead to an 
increase in the effective mass of the bosons which eventually 
gets localized for large interaction strengths . 
But, this happens for integer filling.  Our results (Figs. 4-6)
are for n = 0.25, 0.4, and 1.5 which is not close to integer
filling. The interaction induced enhancement of the boson effective
mass will not be significant in this case since there are
sufficient number of un-occupied sites in the lattice so that
the bosons can move around without paying a penalty for multiple
boson site occupancies. The condensate fraction for bosons in 
various cubic lattices are shown in Fig. 4-6. 
\begin{figure}
\resizebox*{4.5in}{4.7in}{\rotatebox{270}{\includegraphics{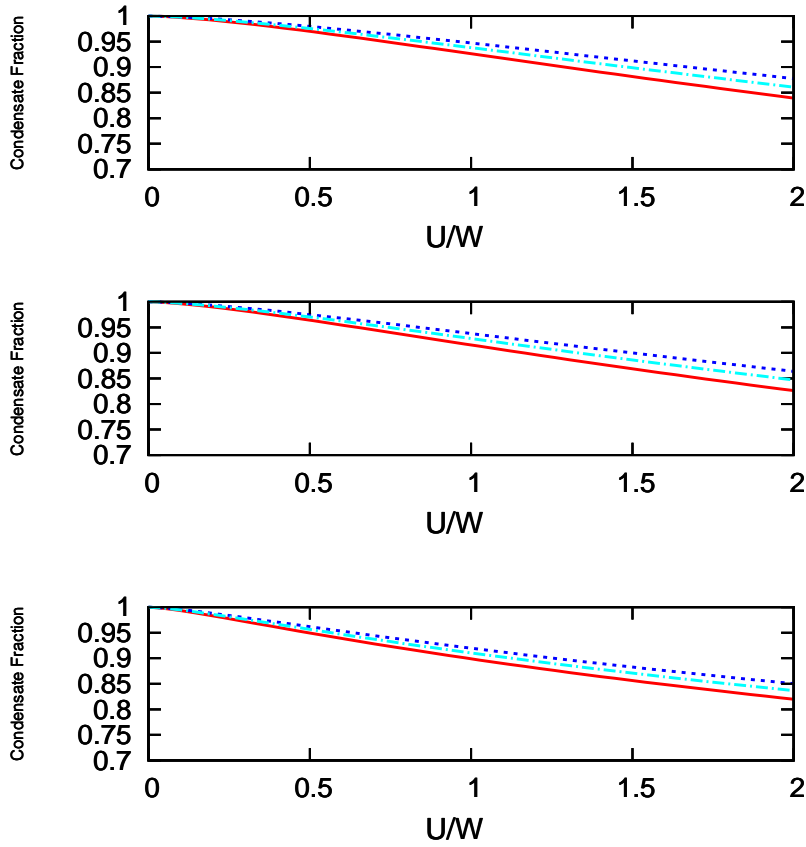}}}
\vspace*{0.5cm}
\vspace*{0.5cm}
\caption[]{
The variation of Condensate fraction (for T=0) with U/W of weakly 
interacting  bosons  in a sc lattice with NNN hopping  $ t^{\prime}$: 
$t^{\prime}$ = 0 (solid line), $t^{\prime}$ = t/20 (dash-dot line), $t^{\prime}$ = t/10 (dotted). Here 
  n =0.25 (top panel), n=0.4 (middle panel), n=1.5 (bottom panel). 
}
\label{scaling}
\end{figure}

\begin{figure}
\resizebox*{4.5in}{4.7in}{\rotatebox{270}{\includegraphics{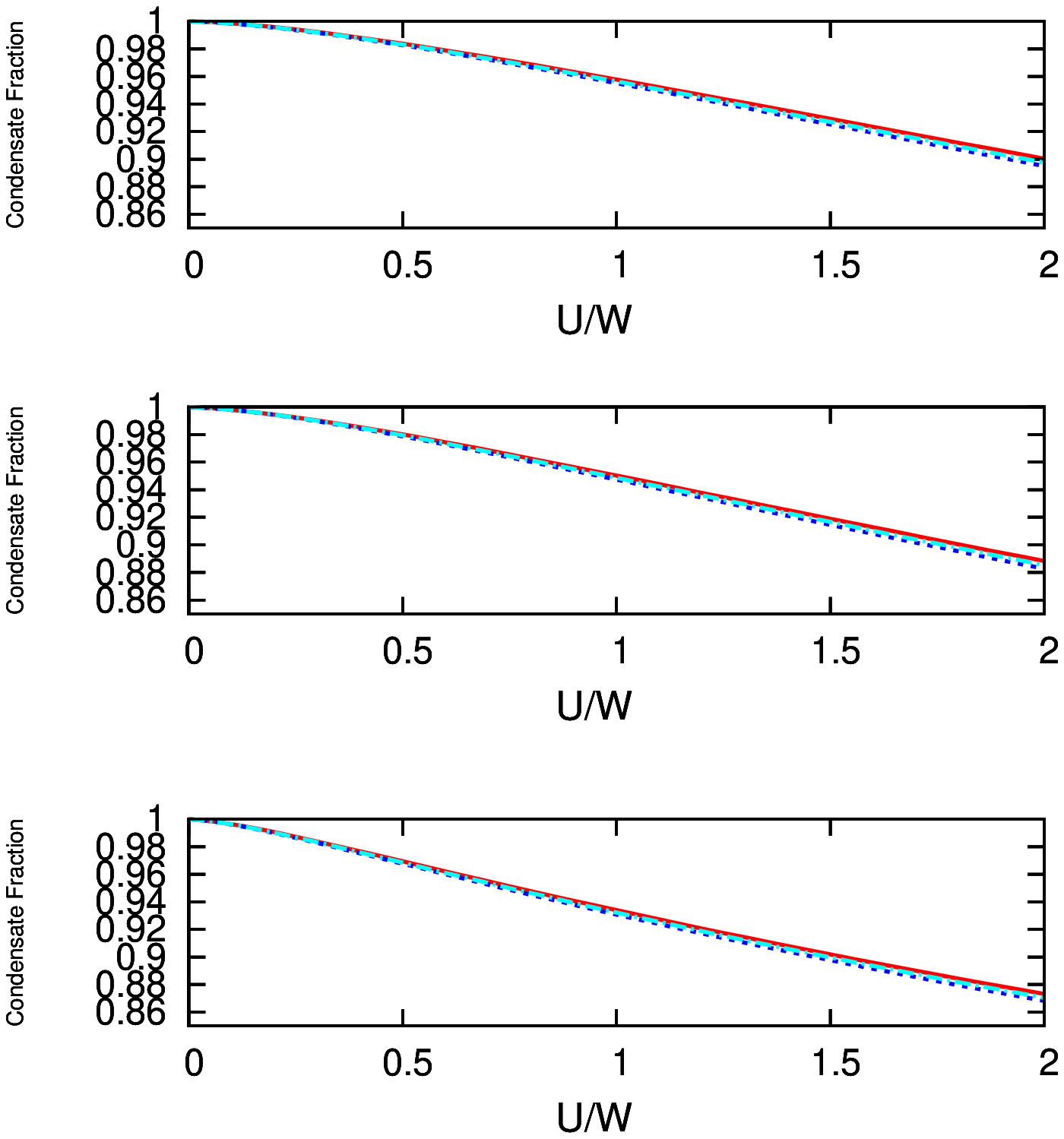}}}
\vspace*{0.5cm}
\vspace*{0.5cm}
\caption[]
{
The variation of Condensate fraction (for T=0 and n=0.4) with U/W of 
weakly interacting  bosons  in a bcc lattice with NNN hopping  
$t^{\prime}$ : $ t^{\prime}$ = 0 (solid line), $t^{\prime}$ = t/20 (dash-dot 
line), $t^{\prime}$ = t/10 (dotted line). Here 
  n =0.25 (top panel), n=0.4 (middle panel), n=1.5 (bottom panel).
}
\label{scaling}
\end{figure}

\begin{figure}
\resizebox*{4.5in}{4.7in}{\rotatebox{270}{\includegraphics{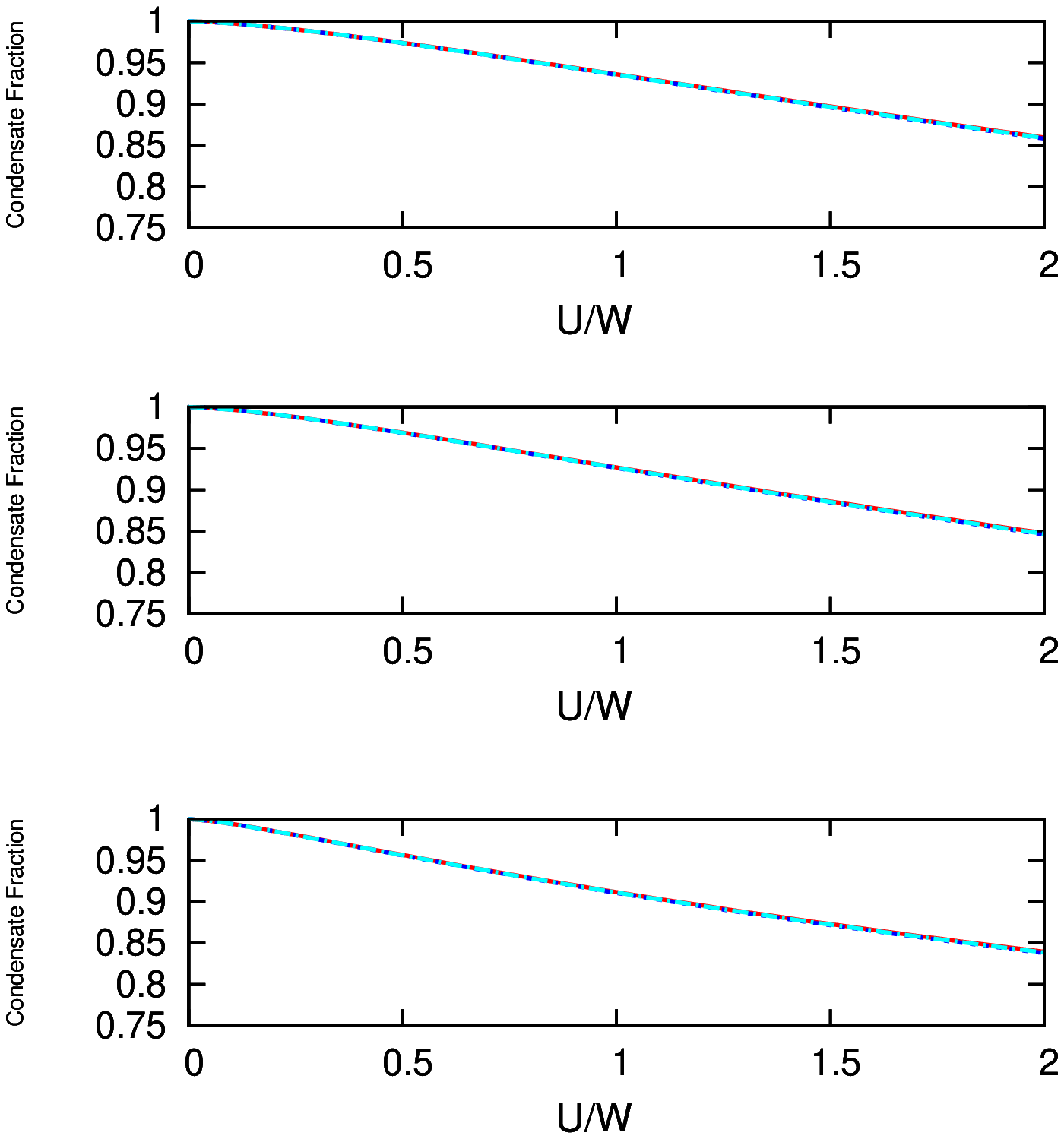}}}
\vspace*{0.5cm}
\vspace*{0.5cm}
\caption[]{
The variation of condensate fraction (for T=0 and n=0.4) with U/W of 
weakly interacting  bosons  in a fcc lattice with NNN hopping  
$t^{\prime}$ : $ t^{\prime}$ = 0 (solid line), $t^{\prime}$ = t/20 (dash-dot 
line), $t^{\prime}$ = t/10 (dotted line). Here 
  n =0.25 (top panel), n=0.4 (middle panel), n=1.5 (bottom panel).
}
\label{scaling}
\end{figure}

For bosons in a sc lattice,
we find that interaction induced depletion of the condensate is reduced
with increasing $t^{\prime}$ as shown in Fig. 4. For bosons in a bcc 
lattice (Fig. 5), increasing $t^{\prime}$ is found increase
the interaction induced depletion. 
In the case of bosons in an fcc lattice, the effects of increasing $t^{\prime}$
does not have much effect on condensate fraction as shown in Fig. 6.
The bose condensation temperature
is unaffected by the interaction in the Bogoliubov method. 
\section{Conclusions}
In this paper, we investigated the effects of NNN hopping
of non-interacting and interacting bosons in cubic lattices on 
bose condensation temperature and ground state occupancy.
We find that the bose condensation
temperature is enhanced with increasing
$t^{\prime}$ for bosons in a simple cubic (sc) lattice and
decreases for bosons in body-centered cubic (bcc) and
face-centered cubic (fcc) lattices. We also find that
interaction induced depletion of the condensate is reduced for bosons
in a sc lattice while it is enhanced for bosons in bcc and fcc lattices. 
These results would be relevant to bosons in condensed matter 
systems in which NNN boson hopping is not negligible. The results
could also be applicable to bosons in optical lattices.
For instance, it was recently shown that hard-core lattice
bosons moving in an optical lattice and interacting with
phonon modes of polar molecules trapped in the lattice develops
significant NNN hopping amplitudes\cite{datyar}. There is a hope that several
models of strongly correlated quantum many particle systems can be
simulated in a controlled manner in optical 
lattice systems\cite{blochdali,yukalov}.
As mentioned earlier, there is also a renewed effort in investigations
of bose condensation in crystalline lattices. Further, one of
the routes to superconductivity is through the condensation of charged bosons
(bipolarons, for example). Furthermore, higher temperature superconductivity
may be possible in correlated electron systems by the
condensation of charged bosons generated within the electrons system through
strong correlation effects\cite{baskaran,senthil}. Though
we are unable to find a concrete example at this juncture, it is not 
inconceivable that the energy spectra of some of the 
possible emergent boson modes in strongly correlated lattice electrons 
systems disperses away in momentum space with significant 
contributions from NNN hopping amplitudes. Then, our results may have 
some relevance to these bosons as well. 
\section{Acknowledgment}
Gopesh Kumar Chaudhary thanks the University Grants Commission (UGC), 
Government of India for providing financial support through a Junior 
Research Fellowship. 


\begin{thebibliography}{99}
\bibitem{neutral}
We note here that the results would also apply to spin-zero neutral
bosons loaded in optical lattices once the interaction term in Eq. 6
is correspondingly interpreted. In both cases (charged or neutral)
the interaction can be written in terms of the scattering lengths.
\end{thebibliography}
\end{document}